%% file: main.tex
\documentclass[sigconf]{acmart}

\renewcommand\footnotetextcopyrightpermission[1]{}
\AtBeginDocument{%
  }

\input{./files/util.tex}

\usepackage{subcaption}
\begin{document}

\title{\tool: A Security Model of AI Agents}

\author{Hongyi Lu}
\email{luhy2017@mail.sustech.edu.cn}
\affiliation{\department{Research Institute of Trustworthy Autonomous Systems}
	\institution{Southern University of Science and Technology}
	\country{China}
}
\additionalaffiliation{
	\institution{Department of Computer Science and Engineering, Southern University of Science and Technology}
}
\additionalaffiliation{
	\institution{Department of Computer Science and Engineering, Hong Kong University of Science and Technology}
}

\author{Nian Liu}
\email{12132347@mail.sustech.edu.cn}
\affiliation{
	\institution{Department of Computer Science and Engineering Southern University of Science and Technology}
	\country{China}
}
\additionalaffiliation{\department{Research Institute of Trustworthy Autonomous Systems}
	\institution{Southern University of Science and Technology}
}

\author{Shuai Wang}
\email{shuaiw@cse.ust.hk}
\affiliation{
	\department{Department of Computer Science and Engineering}
	\institution{Hong Kong University of Science and Technology}
	\country{China}
}

\author{Fengwei Zhang}
\email{zhangfw@sustech.edu.cn}
\authornotemark[3]
\affiliation{
	\department{Department of Computer Science and Engineering}
	\institution{Southern University of Science and Technology}
	\country{China}
}

\begin{abstract}
	Autonomous AI agents powered by Large Language Models can reason, plan, and
	execute complex tasks, but their ability to autonomously retrieve information
	and run code introduces significant security risks. Existing approaches
	attempt to regulate agent behavior through training or prompting, which does
	not offer fundamental security guarantees. We present \tool, a security
	framework that enforces formally verified policies on AI agents under a
	worst-case threat model where the agent itself may be adversarial. \tool
	formalizes a fine-grained security model over system entities, trust scopes,
	and permissions to express dynamic policies that adapt to agents' runtime
	behavior. These policies are translated into concrete security rules and
	enforced through a user-space kernel augmented with BPF-based syscall
	interception. This approach bridges the formal security model with practical
	enforcement, ensuring security regardless of the agent's internal design.
\end{abstract}


\maketitle

\input{./files/intro.tex}
\input{./files/bg.tex}
\input{./files/tm.tex}

\input{./files/design.tex}
\input{./files/discussion.tex}

\bibliographystyle{ACM-Reference-Format}
\bibliography{ref}
\balance

\appendix

\end{document}

%% file: files/util.tex
\newcommand{\tool}{\mbox{ClawLess}\xspace}

\definecolor{wsorange}{RGB}{245,166,115}

\newcommand{\parh}[1]{\noindent\textbf{#1}}

\usepackage{xspace}
\usepackage{amsmath,amsfonts}
\usepackage{graphicx}
\usepackage{textcomp}
\usepackage[x11names]{xcolor}
\usepackage{threeparttable}
\usepackage{colortbl}
\usepackage{graphicx}
\usepackage{listings}
\usepackage{color}
\usepackage{array}
\usepackage{float}
\usepackage{graphicx}
\usepackage{multirow}
\usepackage{colortbl}
\usepackage[ruled,linesnumbered,boxed]{algorithm2e}
\usepackage{algpseudocode}
\usepackage{pifont}
\usepackage{enumitem}
\usepackage{balance}
\usepackage{url}
\usepackage[T1]{fontenc}
\usepackage[utf8]{inputenc}
\usepackage{caption}
\usepackage{booktabs}
\usepackage{tabularx}
\usepackage{diagbox}
\usepackage{amsmath}
\usepackage{listings}
\usepackage{makecell}
\usepackage{verbatim}
\usepackage{acronym}
\usepackage[super]{nth}
\usepackage{tikz}
\usepackage{amsmath}
\usepackage[misc]{ifsym}
\usepackage{grffile}
\usepackage{tikz,pgf}
\usepackage{amsthm}
\usepackage[most]{tcolorbox}
\usepackage{enumitem}
\usepackage{trimclip}

\usetikzlibrary{calc}
\makeatletter
\def\mfontsize{\f@size}

\newcommand{\F}{Fig.}

\newcommand{\T}{Table}
\renewcommand{\S}{Sec.}

\definecolor{codegreen}{rgb}{0,0.6,0}
\definecolor{codegray}{rgb}{0.5,0.5,0.5}
\definecolor{codepurple}{rgb}{0.58,0,0.82}
\definecolor{backcolour}{rgb}{0.95,0.95,0.92}
\definecolor{cadmiumgreen}{rgb}{0.0, 0.42, 0.24}
\lstdefinestyle{mystyle}{
	escapeinside={(*@}{@*)},
	backgroundcolor=\color{backcolour},
	commentstyle=\color{codegreen},
	keywordstyle=\color{magenta},
	numberstyle=\tiny\color{codegray},
	stringstyle=\color{codepurple},
	frame=shadowbox,
	basicstyle=\ttfamily\footnotesize,
	breakatwhitespace=false,
	breaklines=true,
	captionpos=b,
	keepspaces=true,
	numbers=left,
	numbersep=5pt,
	showspaces=false,
	showstringspaces=false,
	showtabs=false,
	tabsize=2,
}

\lstdefinestyle{gdb}
{
	backgroundcolor=\color{backcolour},
	keywordstyle=\color{magenta},
	escapeinside={(*@}{@*)},
	basicstyle=\scriptsize\color{black}\ttfamily,
	frame=shadowbox
}

%% file: files/intro.tex
\section{Introduction}
\label{sec:intro}

Autonomous AI agents represent a paradigm shift in how computing systems
interact with the digital world. Unlike traditional software that follows
predetermined instructions, AI agents leverage Large Language Models (LLMs) to
reason, plan, and execute complex tasks autonomously~\cite{gpt4,claude,openclaw}. Frameworks like
OpenClaw, OpenCode and Claude Code have enabled AI agents to autonomously
conduct a series of actions from information retrieval, code execution and
results validation, forming an end-to-end solution to various tasks. However,
this powerful automation also brings significant security risks~\cite{openclawcves}.

Various cases from the real world have shown that the security risks faced by
autonomous AI agents are not theoretical. Recent studies~\cite{ipi} show that
it is possible to alter AI agents' behavior by publishing malicious content
online, which is retrieved and interpreted by AI agents, leading to
exploitation. Moreover, agents nowadays have significantly expanded their
capability boundaries by being equipped with a diverse set of \textit{tools}.
Studies~\cite{aisupply} show that these tools lack sufficient security auditing,
leading to severe security risks. Additionally, AI agents also face risks from
its infrastructure. \citet{fun} shows that it is possible to hijack
the control flow of a GPU program by memory corruptions.
These incidents highlight the urgent need for a security solution for AI
agents. However, designing such a solution is not a trivial task; the following
challenges exist.

\parh{Ambiguous Trust Boundary.} Unlike traditional software that
retrieves data from a fixed set of endpoints with clear trusted/untrusted
policy, AI agents retrieve data autonomously from diverse sources. This
enables agents to conduct sophisticated tasks, but also blurs the boundary
between trusted and untrusted input; the latter could exploit the agents to
behave maliciously.

\parh{Privilege/Usability Trade-off.} AI agents often require a variety of
privileges to conduct tasks effectively, such as file system access, network
connectivity, and even program execution. However, granting
these privileges creates security risks. Traditional least-privilege
principles are challenging to apply to AI agents. Overly restrictive
privilege models limit the agent's usability, while permissive approaches expose
systems to potential exploitation.

\parh{Security for Autonomous Software.} Though being able to
autonomously retrieve information online and execute commands brings
significant advantages to agents in complex tasks, they also grant
agents the capability to theoretically break any security mechanism. For
example, a malicious prompt online might mislead the agents to escape its
Docker~\cite{docker} container with a detailed description of a 0-day in Docker.

Traditional security mechanisms are inadequate for this threat model. Static
analysis of agent behavior is fundamentally infeasible due to the
non-deterministic nature of LLM outputs. Sandboxing approaches like
Docker~\cite{docker} provide isolation but remain dependent on the underlying
Linux kernel, inheriting its vulnerabilities~\cite{dockercves}. Formal
verification techniques, while powerful, have not been adapted to the dynamic,
tool-using behavior of AI agents.

In this paper, we present \tool, a system designed to secure autonomous AI
agents through a combination of fine-grained privilege analysis, secure
container isolation and runtime auditing. Our key insight is that AI agent
security requires a comprehensive analysis of its security policies and a
trusted foundation to enforce these policies.
Unlike prior work that focuses on securing individual components~\cite{ace,isogpt} or
simply wrapping agents in containers~\cite{ironclaw}, \tool provides a holistic
framework for deploying AI agents under a ``worst-case'' threat model where the
agent itself may be adversarial.

In summary, this paper makes the following contributions:

\begin{itemize}[leftmargin=*]
	\item We present the first comprehensive security analysis for
	      autonomous AI agents with two fundamental assumptions in AI agent's security.

	\item We formalize a fine-grained security model for AI agents that prevents
	      them from abusing their capabilities to compromise the system, while
	      maintaining their usability.

	\item Based on our formalized security model, we design and implement \tool,
	      an isolation framework that enforces formally verified security policies on
	      AI agents, ensuring their security regardless of their internal design and
	      implementation.
\end{itemize}

%% file: files/bg.tex
\definecolor{deepgreen}{RGB}{0,100,0}       
\definecolor{deepyellow}{RGB}{204,102,0}    
\definecolor{deepred}{RGB}{150,0,0}        

\begin{table*}[htbp]
	\centering
	\caption{Comparison between secure containers.}
	\label{tab:compare}
	\resizebox{.9\linewidth}{!}{
		\begin{tabular}{llcccccc}
			\hline
			\textbf{Type}                   & \textbf{Example} & \textbf{HW} & \textbf{Indep. Kernel} & \textbf{Compatibility\footnotemark[1]} & \textbf{Interoperability\footnotemark[2]} & \textbf{Deployability\footnotemark[3]} & \textbf{Security}          \\ \hline
			\textbf{Standard Container}     & Docker           & No          & No                     & \color{deepgreen}{High}                & \color{deepgreen}{High}                   & \color{deepgreen}{High}                & \color{deepred}{Low}       \\ \hline
			\textbf{User-space Kernel}      & gVisor           & No          & Nested                 & \color{deepgreen}{High}                & \color{deepgreen}{High}                   & \color{deepgreen}{High}                & \color{deepgreen}{High}    \\ \hline
			\textbf{Virtualization}         & Kata Container   & VM          & VM                     & \color{deepgreen}{High}                & \color{deepyellow}{Medium}                & \color{deepred}{Low}                   & \color{deepgreen}{High}    \\ \hline
			\textbf{Confidential Container} & CoCo             & TEE         & VM                     & \color{deepred}{Low}                   & \color{deepred}{Low}                      & \color{deepred}{Low}                   & \color{deepgreen}{Highest} \\ \hline
		\end{tabular}
	}
\end{table*}

\section{Background}
\label{sec:bg}

\subsection{Secure Containers}
\label{bg:container}

\begin{figure}[htbp]
	\centering
	\includegraphics[width=.8\columnwidth]{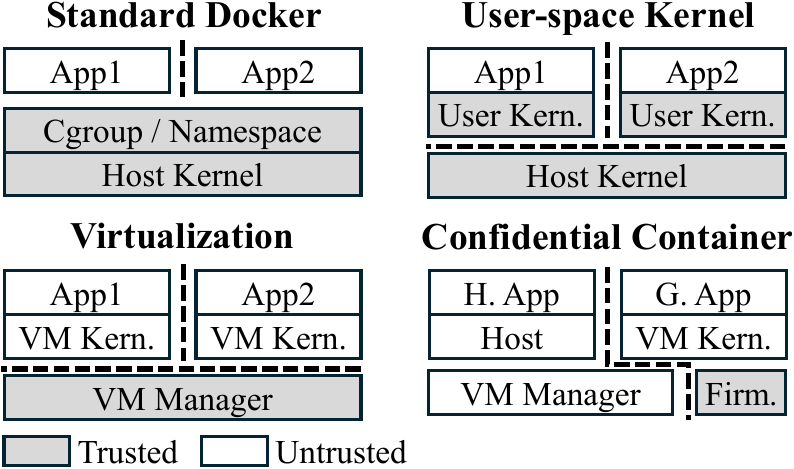}
	\caption{Architectures of secure containers.}
	\label{fig:compare}
\end{figure}

\tool relies on secure containers to isolate potentially malicious AI agents; we
therefore introduce them here. Standard Docker~\cite{docker}, as illustrated in
\F~\ref{fig:compare}, directly
rely on the isolation facilities like \texttt{cgroup} and \texttt{namespace} of
Linux kernel to execute unmodified applications in isolation. However, as Linux
is a monolithic codebase, this approach suffers from various vulnerabilities
occurring in Linux. Another approach for secure containers is to provide a
user-space kernel and execute the application atop. This method deploys a
trustworthy layer~(e.g., gVisor~\cite{gvisor}) between the potentially malicious
application~(e.g., AI agents) and the vulnerable kernel, ensuring the security
of the entire system. Both Docker and user-space kernel are pure software
solutions, virtualization, on the other hand, utilizes hardware extensions
to host an independent VM kernel for the application.
Lastly, confidential containers like CoCo~\cite{coco} take the most drastic
approach of isolating the containers inside a Trusted Execution Environment~(TEE),
excluding the Virtual Machine Manager~(VMM) from the trusted codebase.

The architectural differences among these secure containers also bring different
properties to them. \T~\ref{tab:compare}
presents a comprehensive
comparison of these secure container technologies in terms of compatibility,
interoperability, deployability, and security. From the table, we can see that
despite standard Docker achieves the highest usability by sharing host kernel,
it suffers from weak security isolation due to its reliance on the monolithic
host kernel; it has 37 CVEs over the past ten years and five of them are
high-severity vulnerabilities with a >9.0 CVSS. Containers that utilize user-space
kernel, on the other hand, achieves a balance between usability and security;
it protects the host kernel by delegating most kernel-user interactions to the
user-space kernel~(only one CVE in past ten years) while maintaining usability by
acting as a user-space process on the host. Virtualization and confidential
containers achieve high security assurance by using hardware-based isolation.
However, this hardware boundary also creates obstacles for their interaction
between the guest and host environment, which prevents AI agents from
conducting tasks that involve the host environment. Moreover, both virtualization and
confidential container approaches require high privilege to deploy, which is infeasible in
cloud environments.

\subsection{Berkeley Packet Filter}
\label{bg:monitor}

\begin{figure}[htbp]
	\centering
	\includegraphics[width=.6\columnwidth]{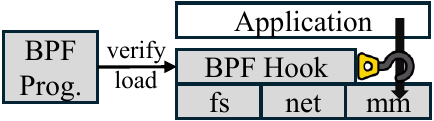}
	\caption{BPF workflow.}
	\label{fig:bpf}
\end{figure}

\tool uses Berkeley Packet Filter~(BPF) to achieve the monitoring of AI
agents' behaviors. BPF is a kernel facility that allows customized programs to
run safely in kernel space without modifying kernel source or loading kernel
modules. As shown in \F~\ref{fig:bpf}, a BPF program
can be attached to a set of kernel hookpoints~(e.g., syscalls). Depending on
the hookpoint, the program can conduct tasks from syscall tracing to packet
filtering.

BPF offers several key advantages for security monitoring: it compiles to
native code, executes with minimal overhead and supports dynamic policy updates
without kernel recompilation. In sum, BPF provides fine-grained visibility into
system events (e.g., syscalls, network operations, file accesses).

%% file: files/tm.tex
\newtcolorbox{boxK}{
	boxsep=0pt,
	sharpish corners, 
	boxrule = 0pt,
	toprule = 0pt, 
	enhanced,
	fuzzy shadow = {0pt}{-2pt}{-0.5pt}{0.5pt}{black!35}
}

\section{Threat Model}
\label{sec:tm}

Unlike traditional software which executes a series of predefined instructions,
AI agents are able to autonomously learn information from outside world and
execute commands. This not only grants AI agents the capabilities of solving
complicated tasks, but also creates a severe security risk: if AI agents can
accomplish these complicated tasks, how to ensure they cannot break our security
mechanisms. We establish the assumptions of AI agents' capabilities.

\begin{boxK}
	\textit{Assumption of Capabilities}: AI agents are capable of conducting sophisticated attacks against any
	security mechanisms.
\end{boxK}

Worse still, our deployment and application of AI agents now have
largely exceeded our understanding of them. Though extensive efforts have been
put into researching the theory of machine learning, there is \textit{not} yet
a solid foundation of machine learning. As there are no theoretical guarantees over
agents' behavior and AI agents are constantly exposed to unsanitized input online,
we can reach the following assumption.

\begin{boxK} \textit{Assumption of Maliciousness}: AI agents will eventually be
	lured to become malicious. \end{boxK}

\parh{Trusted components.}~We assume the software~(e.g., gVisor) and the
hardware used to implement \tool are free of vulnerabilities and
thereby trustworthy. Though such software/hardware does not exist in practice,
\tool realizes this assumption in a best-effort manner by following the
principle of least privilege and shrinking the attack surface with minimized
trusted codebase.

\parh{Protected components.}~This includes the host software stack, host kernel
and components that are not used for security purposes. We assume these
components do not act maliciously, but they could be compromised
due to the vulnerabilities presented in their code. Thus, these components
are protected from untrusted ones.

\parh{Untrusted components.}~The remainder of the system, such as the AI agent
itself and its software stack inside the container are regarded as untrusted.
This is because AI agents process a large amount of unsanitized input, which
could alter its behavior and make it become malicious. We therefore isolate
these components from the rest of the system.

\parh{External components.}~This refers to the external components~(e.g.,
script/code) downloaded or generated by the AI agents. Unlike traditional
software, AI agents process a diverse collection of data; some of them are even
executable programs. These components are from unsanitized sources and more
likely to contain malicious payloads. We thereby further establish additional
isolation to safely utilize these components. Though this additional isolation
might seem unnecessary given our assumption that AI agents eventually
become malicious, we wish to stall this in a best-effort manner to
maintain their usability.

\parh{Out-of-scope.}~As stated in \textit{Assumption of Maliciousness}, AI
agents eventually become malicious due to their exposure to unsanitized
information, it is natural for \tool to exclude DoS attacks. That said, we
still deploy mechanisms such as sandbox to stall such attacks. As we focus on
the security of AI agents, we do not consider vulnerabilities occurring in the
software stack utilized by AI agents~(e.g., vulnerabilities in NodeJS).
Similarly, we also exclude micro-architecture attacks such as Spectre and
Meltdown.

%% file: files/design.tex
\section{Design}
\label{sec:design}

\subsection{Motivation and Overview}
\label{design:overview}

\begin{figure}[H]
	\centering
	\includegraphics[width=0.7\columnwidth]{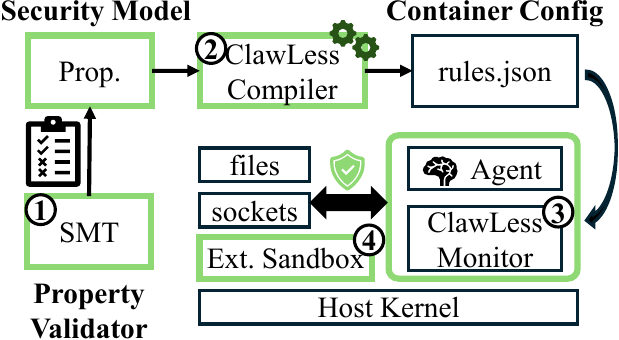}
	\caption{Overview of \tool.}
	\label{fig:overview}
\end{figure}

\parh{Motivation.}~The design of \tool is guided by a fundamental insight: \textit{AI agents
	fundamentally break the threat models underlying traditional security
	mechanisms}. Conventional software security assumes static trust boundaries,
applications retrieve data from predetermined endpoints, execute fixed sets of
instruction sequences, and operate within analyzable privilege scopes. AI
agents violate all three assumptions: they autonomously retrieve data from
arbitrary sources, exhibit non-deterministic behavior via LLM inference, and
can reason about security mechanisms to subvert them.
This insight reveals that securing AI agents requires rethinking how security
policies span multiple abstraction levels.
First, policies must be specified at a semantic level independent of
implementation details, enabling security guarantees that remain valid
regardless of how the agent internally reasons or acts. Second, the enforcement
architecture must directly integrate policy specification with concrete
enforcement mechanisms, ensuring that high-level security properties are
preserved when translated to low-level system operations. Third, multiple trust
domains must coexist under a unified semantic model, allowing fine-grained
mediation between external scripts, the agent, and host resources.

\parh{Overview.}~\tool achieves these goals by building a unified framework
across multiple security mechanisms. As depicted in \F~\ref{fig:overview}
, \tool consists of two major parts: formal
parts and concrete parts. The formal part includes a SMT-based validator and a
policy compiler; they validate and translate abstract formal models into
concrete security configurations (\ding{192}\ding{193} in
\F~\ref{fig:overview}). The concrete part
includes a combination of mechanisms like user-space kernel, runtime monitoring
and external sandboxes to ensure the AI agents comply with the security model
generated~(\ding{194}\ding{195} in
\F~\ref{fig:overview}).

In this design, we successfully express and validate the security model of
\tool in a formal language with \ding{192} and \ding{193}. We also bridge this
formal model with concrete sandbox strategies and enforce it in \ding{194} and
\ding{195}, effectively addressing the challenges mentioned before.
The remainder of this section is organized as follows. \S~\ref{design:model}
introduces how we formalize the security model
of AI agents. \S~\ref{design:policies}
discusses \tool's security
policies for common AI agent tasks. Lastly, \S~\ref{design:impl}
shows how \tool translates the formal
models into concrete security configurations, and enforce them with user-space
sandbox and runtime monitoring.

\subsection{Formal Model} \label{design:model} The above
assumptions draw a holistic view of \tool. We now formalize the security model
of \tool in a fine-grained manner.

\parh{Entities.}~We begin by formalizing different entities in a system.
Formally, we define the following set of entities, which is a union of sets
with different types of entities~(e.g., ${E}_\text{proc}$ denotes the set of
processes on the system).
\[{E}={E}_\text{file}\sqcup{E}_\text{proc}\sqcup{E}_\text{socket}\sqcup{E}_\text{dev}\sqcup\cdots\]
\parh{Scope.}~In addition to entities, we further define three scopes of
isolation. Formally, a scope $s\in S=\{Sandbox,Agent,Monitor\}$, decides if a
security policy applies to sandboxed external components~(e.g., script
downloaded by agent), the agent itself or the \tool's monitor~(e.g., gVisor).
The scope set could be extended to other environments such as TEEs (e.g., for
managing credentials), as long as there is no overlapping in their semantics.

\parh{Permission.}~We define the following set of permissions for entities in
the system: \[P=\{Read,Write,NoExecute,Append,Visible\}\] These permissions
govern the operations that a scope may perform on an entity. Among them,
$Read$, $Write$, $Append$, and $Visible$ are permissions that define what
operations a scope can perform on an object, meaning a more trusted domain
is able to operate more permissively. However, $Execute$ indicates a
fundamentally different semantic, it delegates the current execution
environment to a foreign entity; the more trusted a scope is, the \textit{less
	permissive} it should be able to execute another entity. Therefore, we use its
inverse $NoExecute$ to have a consistent semantics across these permissions:
the more trusted a scope is, the more permissions it shall have, including the
restrictive permissions like $NoExecute$. Moreover, we also define two
additional permissions, $Append$ and $Visible$; the former indicates the entity
can only be appended but not destructively overwritten, while the latter
introduces a new semantic that an entity can be referenced in the current scope
without exposing its content. This prevents the content of the credentials
from being leaked into agents' context when they use credentials to access external
services.

\begin{table}[htbp]
	\resizebox{\columnwidth}{!}{
		\begin{tabular}{cccccc}
			\hline
			\textbf{}        & \textbf{File}                                 & \textbf{Directory}                                      & \textbf{Process}                                                                    & \textbf{Socket} & \textbf{Device}                                                             \\ \hline
			\textbf{Read}    & Read                                          & List                                                    & \multirow{2}{*}{\begin{tabular}[c]{@{}c@{}}\\[-0.5em]IPC\\ primitives\end{tabular}} & Receive         & \multirow{4}{*}{\begin{tabular}[c]{@{}c@{}}Device\\ dependent\end{tabular}} \\ \cline{1-3} \cline{5-5}
			\textbf{Write}   & Write                                         & \begin{tabular}[c]{@{}c@{}}Create\\ Delete\end{tabular} &                                                                                     & Send            &                                                                             \\ \cline{1-5}
			\textbf{Append}  & Append                                        & Create only                                             & /                                                                                   & /               &                                                                             \\ \cline{1-5}
			\textbf{NoExec}  & No Exec.                                      & /                                                       & No Fork                                                                             & /               &                                                                             \\ \hline
			\textbf{Visible} & \multicolumn{5}{c}{View/reference the entity}                                                                                                                                                                                                                                                 \\ \hline
		\end{tabular}}
\end{table}

\parh{Attributes.}~In addition to permissions, \tool defines various
attributes of an entity for practical policy specification. For example, file
path is an attribute of file entities (e.g.,
$\text{Path}(E)=\texttt{/etc/passwd}$). Formally, an attribute is a map from
entity set to its attribute range $A: S\rightarrow V_A$, where $A$ is either an integer
set~(e.g., PID) or string set defined by a regular expression~(e.g., \texttt{/root/*}). These attributes allow us to
define rules for specific entities on the system. For example, to isolate
files in path \texttt{/secure} from the agent, a policy can be defined formally
as follows.
\[\forall \text{Path}(e)\in \{\texttt{/secure/*}\}\Rightarrow P(e,Agent)=\emptyset\]

\parh{AI agents.}~Lastly, we formalize the system of an AI agent as a set
$C=(E,A,P)$, which consists of entities and their corresponding
attributes/properties available to the AI agents. This set $C$ represents the
resources that are accessible to the AI agent. An extra set of security
propositions followed by AI agents $\Sigma$ can be augmented to $C$ to form a
complete security model. For brevity, we use $e\in C$ to denote that the
content of $e$ is in the context of agents.

\parh{Linear Temporal Logic.}~To now, \tool's security model is still static,
meaning it cannot dynamically alter its policies depending on the previous
actions of the AI agent. This significantly restricts the usability of AI agents
as they could be imposed with unnecessary restrictions, especially under our
assumptions of maliciousness. For example, in the static security model, if an
AI agent is simultaneously granted the permission to read a sensitive
file and the permission to write to an out-bound socket. This
automatically implies that the content of that sensitive file is leaked. This
is because, unlike traditional software with predetermined information flow, AI
agents are essentially black-boxes without predictable patterns. In this
particular case, an AI agent can simply read the sensitive file $e_f$ and then
recite its content into the web socket $e_s$. Formally, we have:
\begin{align*}
	Read\in P(e_f,Agent)                    & \Rightarrow e_f\in C    \\
	Write\in P(e_s, Agent)  \wedge e_f\in C & \Rightarrow Leaked(e_f)
\end{align*}
To avoid such malicious behaviors in a static model, the system must be
restricted with the policies that either all sensitive files are not readable
to the agent, or the agent has no out-bound sockets, ignoring the fact the
agent might not have read the sensitive file in the first place. To address
this issue, we augment permissions $p$ with a temporal argument $t$~(i.e., $p$
is invoked at time $t$) and a set of temporal predicates
$\{\square$\footnote{$\square P$ means $P$ holds at all future
	states.},$\lozenge$\footnote{$\lozenge P$ means $P$ holds at some future
	states.}$\}$. We can then define the following rules to prevent out-bound sockets
only after the agents have read the content of sensitive files.
\[Read\in P(e_f,Agent,t)\Rightarrow \square \lnot Write \in P(e_s,Agent)\]

\subsection{\tool's Security Policies}
\label{design:policies}

\parh{Scope Hierarchy.}~One of the most fundamental policies in \tool is the
policy of \textit{Scope Hierarchy}. It dictates a simple fact: the permissions
in a more restricted scope~(e.g., sandbox) should not exceed the permissions of
its parents~(e.g., agents). Formally, this can be expressed in the following
proposition. \[\forall e\in E, P(e,Sandbox)\subseteq P(e,Agent)\subseteq P(e,
	Monitor)\] This proposition specifies that the permissions of the same entity
$e$ in the sandbox scope must be less or equal to the permissions in the agent
scope. Similarly, the permissions in the agent scope must also be less or equal
to the permissions of the monitor. This policy, though seemingly mundane, can
detect insecure policies generated by misconfigured models.

\parh{External Script Sandbox.}~To meet the requirements of diverse tasks, it
is common for AI agents to download external scripts and execute them locally.
These unsanitized scripts often contain harmful content and lead to various
security consequences. To mitigate this, the following policy can be defined to
enforce all scripts that are agent-executable to be executed in sandbox.
\[\forall e\in E, NoExec\in P(e,Agent)\]

\parh{Credential Visibility.}~Based on our assumption over agents'
maliciousness, none of the important credentials should enter the context of
agents. However, to fully unleash the potential of AI agents, they still need
these credentials to access external services. To address this issue, we define
the permission of $Visible$ back in \S~\ref{design:model}
to allow the agent to reference a credential
without accessing its content. Formally, we can define the following policy to
enforce that all credentials are solely visible to the agent but not readable or writable.
\[\forall e_f,Credential(e_f)\Rightarrow P(e_f,Agent)=Visible\]

\parh{SMT-based Policy Validation.}~All these policies can be expressed
formally using the model we defined in
\S~\ref{design:model}. This further
enables us to use SMT solvers like z3~\cite{z3} to validate if the security
policies are satisfied by a given configuration. For example, if a
developer grants the agent the permission to execute a script, the SMT
solver can immediately detect this is a violation of the external script
sandbox policy and alert the developer.

\begin{table*}[htbp]
	\begin{tabular}{cccccc}
		\hline
		\multicolumn{1}{l}{} & \textbf{Read}                               & \textbf{Write}          & \textbf{Append} & \textbf{NoExec} & \textbf{Visible}      \\ \hline
		\textbf{File}        & read/mmap/sendfile/...                      & write/mmap/sendfile/... & lseek/open/...  & execve          & \multirow{4}{*}{stat} \\ \cline{1-5}
		\textbf{Directory}   & getdents                                    & mkdir/rmdir/creat/...   & mkdir/creat/... & /               &                       \\ \cline{1-5}
		\textbf{Socket}      & recvfrom/recvmsg/...                        & sendto/sendmsg/...      & /               & /               &                       \\ \cline{1-5}
		\textbf{Device}      & \multicolumn{3}{c}{ioctl/iopc/...}          & /                       &                                                           \\ \hline
		\textbf{Process}     & \multicolumn{3}{c}{semget/semop/semctl/...} & clone/fork              & proc fs                                                   \\ \hline
	\end{tabular}
	\caption{Mapping from \tool's security model to system calls.}
	\label{tab:syscall_mapping}
\end{table*}

\subsection{Policy Enforcement} \label{design:impl} With
the formal model and policies defined in the previous sections, we now
discuss how \tool translates these formal models into concrete security
rules and enforce them on the AI agents.

\parh{Policy Compiler.}~Despite its autonomous nature, an AI agent still relies
on the ubiquitous \textit{system call} interface~\cite{syscall} for tasks like file
access, network communication and process management. This makes system calls a
critical interface for enforcing security policies on AI agents. However, there
exists a significant gap between the high-level formal model and the low-level
system call interface. For example, the formal model abstractly categorizes
files as entities and defines permissions like $Read$ and $Write$, while the
system call interface defines a series of different system calls like
\texttt{open}, \texttt{read}, and so on. For legacy reasons, two system calls
with almost identical functionality could exist~(e.g., \texttt{open} and
\texttt{openat}). This significantly complicates the enforcement of \tool's
security policies. Fortunately, the system call interface has evolved for
many years and is very stable now. Therefore, this enables us to build a
comprehensive mapping from each of the system calls to the permissions defined
in \tool's formal model.

\T~\ref{tab:syscall_mapping}
shows a mapping
from \tool's security model to Linux's system calls. Due to the semantic gap
between the formal model and the system call interface, certain system calls
might require different permissions depending on their arguments. For example,
the \texttt{sendfile} syscall directly copies data from a file to another file
or socket, which simultaneously requires both $Read$ permission from the source
and $Write$ permission to the destination.

\parh{Syscall interception.}~Once the policies are translated into concrete
syscall rules, \tool deploys a user-space kernel as its monitor to enforce
these rules. Specifically, \tool's monitor utilizes BPF programs to intercept
system calls from the agent and check if they comply with the policies.
Specifically, \tool's monitor uses \texttt{raw\_tracepoint} to intercept
\texttt{sys\_enter} event, which is triggered whenever a system call is
invoked.
\begin{figure}[htbp]
	\centering
	\includegraphics[width=0.8\columnwidth]{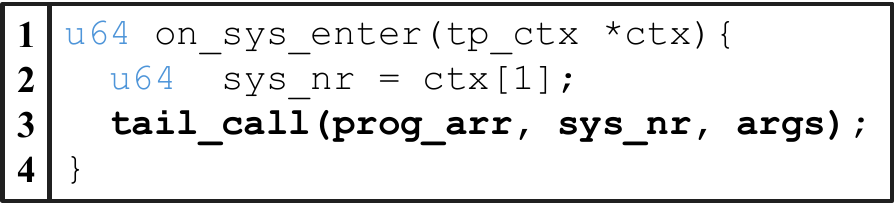}
	\caption{Syscall interception.}
	\label{fig:syscall}
\end{figure}
As shown in \F~\ref{fig:syscall},
the BPF program identifies the system call number~(line 2), and then yields the
control to the designated handler via \texttt{bpf\_tail\_call}~(line
3). Since BPF allows dynamically loading and updating the program in
\texttt{prog\_arr}, \tool can easily update the policies by updating the
corresponding handler without interrupting the system.
\begin{figure}[htbp]
	\centering
	\includegraphics[width=0.8\columnwidth]{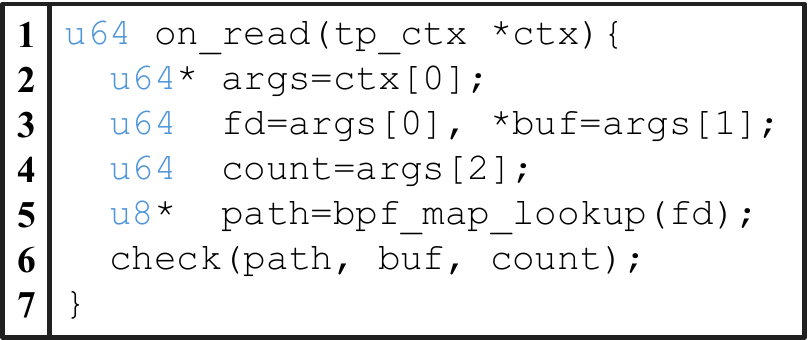}
	\caption{Handler for syscall \texttt{read}.}
	\label{fig:filter}
\end{figure}
\F~\ref{fig:filter} shows an
example handler for the \texttt{read} syscall. The handler first retrieves
arguments of the syscall~(line 2-4), and obtains the corresponding path
information related to the file descriptor~(line 5). With all the necessary
information, the handler can then check if this syscall invocation complies
with the policies defined previously.

\parh{External Sandbox.}~AI agents often need to execute external scripts for
tasks like data processing and code generation. These scripts, however, are
often potentially malicious and can lead to various security consequences. To
mitigate this issue, \tool additionally creates a sandbox environment for these
external scripts. Specifically, \tool utilizes the same user-space kernel
technology to create a sandbox environment with lesser permissions~(see
\S~\ref{design:policies}). In this
way, AI agents can arbitrarily execute external scripts within a clear
permission scope~(i.e., a subset of its permissions).

%% file: files/discussion.tex
\section{Discussion}
\label{sec:disc}

\parh{Prior work.}~With the rise of AI agents like OpenClaw~\cite{openclaw,ironclaw}, there have
been a number of works focusing on improving the security of these powerful yet
potentially insecure agents. IsolateGPT~\cite{isogpt} proposes a framework to isolate
the natural language interactions between different LLM-based applications.
ACE~\cite{ace} abstracts the execution of AI agents into a verifiable execution
plan to ensure the integrity of the data on the system. In addition to
enforcing isolation on these AI agents, there are also many works focusing on
regulating the behavior of these agents by training or prompting
techniques~\cite{antidote,expguard,neurofilter,proact,adasteer}. The lack of
understanding of the internal of these AI agents has made it practically
impossible to fundamentally regulate their behavior. In contrast, \tool takes a
different approach by enforcing security models defined by formal
specifications to ensure the security of these non-deterministic agents,
ensuring their security regardless of their internal design and implementation.

\section{Conclusion}
\label{sec:conclusion}

In this paper, we presented \tool, a security framework designed to protect
systems from potentially malicious autonomous AI agents. Our work is grounded
in two fundamental assumptions: \ding{192} AI agents are capable of
sophisticated attacks, and \ding{193} they will eventually be lured into
malicious behavior, which together demand a security solution that does not rely
on the agent's cooperation.

We formalized a fine-grained security model that captures entities, scopes, and
permissions across multiple domains in the system. Building on this formal
foundation, we designed a policy compiler that translates high-level security
specifications into concrete system call rules, and enforced them through a
user-space kernel augmented with BPF-based syscall interception.

Unlike prior approaches that attempt to regulate agent behavior through
training or prompting, \tool enforces externally specified, formally verified
policies that remain effective regardless of the agent's internal design or
implementation. We believe this approach provides a principled foundation for
securing increasingly capable autonomous AI agents.

\section*{Acknowledgments}

We would like to thank the COMPASS members for their insightful comments. This
work is partly supported by the National Natural Science Foundation of China
under Grant No. U2541211, No. 62372218, and No. U24A6009. This work was also
supported in part by a grant from the Research Grants Council of the Hong Kong
Special Administrative Region, China HKUST C6004-25G and an ITF grant under the
contract ITS/161/24FP.